\documentclass[superscriptaddress,showpacs,amsmath,amssymb,aps,prd,twocolumn,floatfix,nofootinbib]{revtex4-1}

\usepackage{graphicx}
\graphicspath{{figures/}}

\usepackage{dcolumn}
\usepackage{bm}
\usepackage{color}
\usepackage{amsmath}
\usepackage{amssymb}
\usepackage{bbm}
\usepackage{lineno}
\usepackage{hyperref}
\usepackage{todonotes}
\usepackage{hhline}
\usepackage{layouts}
\hypersetup{
    colorlinks = true
}

\usepackage{natbib}
\usepackage{graphicx}
\usepackage{amssymb}
\usepackage{siunitx}
\usepackage{framed}
\usepackage{physics}
\usepackage{siunitx}
\usepackage{todonotes}
\usepackage{lineno}
\newcommand{\outline}[2][]{\todo[disable,inline,color=green!20,#1]{#2}}

\newcommand{\comment}[2][]{\todo[disable,inline,color=blue!20,#1]{#2}}

\begin{document}
\title{Implicit Neural Representation as a Differentiable Surrogate for Photon Propagation in a Monolithic Neutrino Detector}


\newcommand{\SLAC}{SLAC National Accelerator Laboratory, Menlo Park, CA, 94025, USA}
\affiliation{\SLAC}
\newcommand{\Stanford}{Stanford University, Stanford, CA, 94305, USA}
\affiliation{\Stanford}
\newcommand{\Lambdalab}{Lambdalab Inc., San Francisco, CA, 94107, USA}
\affiliation{\Lambdalab}
\newcommand{\Berkeley}{University of California, Berkeley, CA, 94720, USA}
\affiliation{\Berkeley}

\author{Minjie~Lei} \email{minjilei@stanford.edu} \affiliation{\Stanford}
\author{Ka~Vang~Tsang} \email{kvtsang@slac.stanford.edu} \affiliation{\SLAC}
\author{Sean~Gasiorowski} \affiliation{\SLAC}
\author{Chuan~Li} \affiliation{\Lambdalab}
\author{Youssef~Nashed} \affiliation{\SLAC}
\author{Gianluca~Petrillo} \affiliation{\SLAC}
\author{Olivia~Piazza} \affiliation{\Berkeley}
\author{Daniel~Ratner} \affiliation{\SLAC}
\author{Kazuhiro~Terao} \affiliation{\SLAC}

\collaboration{on behalf of the DeepLearnPhysics Collaboration}\noaffiliation


\begin{abstract} 
  Optical photons are used as signal in a wide variety of particle detectors.
  Modern neutrino experiments employ hundreds to tens of thousands of photon
  detectors to observe signal from millions to billions of scintillation
  photons produced from energy deposition of charged particles.  These neutrino
  detectors are typically large, containing $\mathcal{O}(10^2~-~10^5)$ tons of
  target volume, and may consist of many materials with different optical
  properties. As a result, modeling individual photon propagation requires
  prohibitive computational resources. As an alternative to tracking individual
  photons, the experimental community has traditionally used a {\it look-up
  table}, which contains a mean probability of observing a photon per photon
  detector at each grid location in a uniformly voxelized detector volume.
  However, since the size of a table increases with detector volume for a fixed
  resolution, this method scales poorly for future larger detectors.
  Alternative approaches such as fitting a polynomial to the model could
  address the memory issue, but results in poorer performance.  Furthermore,
  both look-up table and fitting approaches are prone to discrepancies between
  the detector simulation and the real-world detector response. We propose a
  new approach using SIREN, a implicit neural representation with periodic
  activation functions. In our approach, SIREN is used to model the look-up
  table as a ``3D scene'' and reproduces the acceptance map with high accuracy.
  The number of parameters in our SIREN model is orders of magnitude smaller
  than the number of voxels in the look-up table. As it models an underlying
  functional shape, SIREN is scalable to a larger detector. Furthermore, SIREN
  can successfully learn the spatial gradients of the photon library, providing
  additional information for downstream applications.  Finally, as SIREN is a
  neural network representation, it is differentiable with respect to its
  parameters, and therefore tunable via gradient descent. We demonstrate the
  potential of optimizing SIREN directly on real data, which mitigates the
  concern of data vs. simulation discrepancies.  We further present an
  application for data reconstruction where SIREN is used to form a likelihood
  function for photon statistics.  
\end{abstract}


\maketitle


\section{Introduction}
\outline{ 
  Need of optical detectors: timing and energy measurement (or focus on LArTPC?
  then timing primarily), association with other detectors (e.g.  TPC)
}

Liquid Argon Time Projection Chambers (LArTPC) are the detector technology of
choice across the Department of Energy's flagship accelerator-based neutrino
experiments, including the Short Baseline Neutrino (SBN) program and the Deep
Underground Neutrino Experiment
(DUNE)~\cite{acciarri2015proposal,abud2021deep}. LArTPCs provide two detection
modalities: charge and light. Neutrino interactions with Ar nuclei produce
secondary particles; the charged particles ionize Ar atoms along their
trajectories to produce electrons and scintillation photons. The electrons
drift in an applied electric field and are recorded by a grid of detection
wires~\cite{acciarri2017design} or pixels~\cite{dwyer2018larpix}. The 3D
position is reconstructed from one- or two-dimensional spatial measurements
combined with the measured time. Scintillation photons travel isotropically and
are measured by optical detectors. The electron signal is spatially precise but
temporally coarse ($\order{\si{\ms}}$) while the photon signal is spatially
coarse and temporally precise ($\order{\si{\ns}}$).

\outline{
  Challenges: the number of photons, detector size, geometry/material complexity
  Traditional approach: look-up table, polynomial fit
  Solution that is scalable and optimizable against real data
}

An important challenge in creating a full LArTPC simulation is the modeling of
{\em optical visibility}, i.e. the probability of observing a photon produced at a
given location in the detector volume.  Optical visibility is estimated by
generating a large number ($\order{10^4-10^6}$) of photons at a single
location, propagating the generated photons through the detector volume, and
recording the number of photons detected from the optical detectors.  The
current procedure is to create a lookup table called the {\em photon library},
where each step in estimating the optical visibility is repeated over the
entire detector at equally spaced points of $\order{\si{\cm}}$ in each
direction.

\comment{[CLOSED]
  Yifan: Maybe mention the concept of "voxel" when you first introduce the
  "photon library"?  \\
  Patrick: Isn't voxel a proper English word?
}

The generation of the photon library is slow and only redone if there is a
change in the underlying detector properties. From a recent study of the ICARUS
detector~\cite{rubbia2011icarus}, currently the world's largest LArTPC in
operation, it took about a week to generate a photon library of about 2 million
sampling points with a $5 \times 5 \times 5~\si{\cm}^3$ voxel size.  While the
generated ICARUS photon library may stay fixed for downstream usage, it is
already limited by the spatial resolution in physics modeling due to its memory
footprint.  It is therefore not feasible to apply the same strategy for larger
detectors such as DUNE-FD~\cite{abi2020tdr4} ($\sim$100$\times$ ICARUS).
Analytical approximations that require less memory have been proposed as
alternatives to the photon library, but it is often challenging to explicitly
express the underlying distribution using common functions. Furthermore,
neither the photon library nor the analytical approximation are amenable to in
situ calibration with detector data due to the slow regeneration time of the
photon library.  Therefore, even if these models could be made computationally
efficient, they would still introduce biases from differences between
simulation and data.

\comment{[FIXED]
    Sean: Why can you not do something like tune an analytical model? Also,
    presumably if these were computationally efficient, you could do some
    fitting -- I would be more specific than computational complexity here
    (``due to the slow regeneration time'', etc) \\
    Patrick: it's difficult to parameterize the visibility in common 
    functions. It is computationally efficient but not the optimal approximation. \\
    Changed as suggested.
}
\comment{[FIXED]
  Yifan: Is it fair to say "photon library" or "analytic models" are difficult
  to propagate through for systematics? As one is difficult to modify and the
  other is difficult to adjust to the calibration?  \\
  Yifan/group: make clearer that SIREN would be base model and then tune-able
  vs regenerate from scratch every time. Tied to -- make clearer how SIREN
  improves on inference \\
  Patrick: it is not related to the systematic propagation. Also see Sean's 
  comment below.
}

This paper will address these challenges by constructing a differentiable
optical visibility from a neural implicit model. It is designed to learn the
average photon yield at an optical detector from a given a continuous 3D
position inside the detector Such a model is easily and quickly tunable via
gradient descent, providing an efficient method of in situ calibration. As it
is a continuous function, it drastically reduces the number of parameters
needed for physics modeling relative to a voxel representation, and
correspondingly has a much smaller memory footprint, allowing scalable modeling
of optical information in large detectors.

\comment{[FIXED]
  Sean: maybe add more specifics of what SIREN will do here, e.g. ``Such a
  model is easily and quickly tunable via gradient descent, providing an
  efficient method of in situ calibration, and drastically reduces the number of
  parameters needed for physics modeling, reducing the required memory footprint
  and thus allowing the use of optical information in modeling large detectors'' \\
  Patrick: Good suggestion. Added to the text.
}

\section{SIREN for Photon Propagation}
\outline{What is and Why SIREN}

Implicit neural representations are a novel way to parameterize signals as {\em
continuous} functions via neural networks, which are trained to map the domain
of the signal (e.g. spatial coordinates) to the target outputs (e.g. the
signal at those coordinates). Recent advancements in neural scene
representation have demonstrated that neural implicit models can represent 2D
and 3D images containing high frequency features with the level of precision
required by high energy
physics~\cite{mildenhall2020nerf,martel2021acron,sitzmann2020siren}.

Sinusoidal representation networks (SIRENs)~\cite{sitzmann2020siren} are
implicit neural representations that use a simple multilayer perception (MLP)
network architecture along with periodic sine function activations, i.e., 
\begin{eqnarray}
  \vb{\Phi}(\vb{x}) &=& \vb{W}_n(\phi_{n-1} \circ \phi_{n-2} \circ \dots 
  \circ \phi_0) (\vb{x}) + \vb{b}_n, \nonumber \\
  \phi_i(\vb{x}_i) &=& \sin(\vb{W}_i\vb{x}_i + \vb{b}_i),
\end{eqnarray}
where the function $\phi_i : \mathbb{R}^{M_i} \mapsto \mathbb{R}^{N_i}$
($N_i,M_i$ are positive integers) of the $i$-th layer of the network
consists of an affine transform on the input $\vb{x}_i \in \mathbb{R}^{M_i}$
given by the weight matrix $\vb{W}_i \in \mathbb{R}^{N_i \times M_i}$ and the
biases $\vb{b_i} \in \mathbb{R}^{N_i}$, followed by the sine activation applied
on each component of the resulting vector. The input signal is parameterized as
a continous function $\vb{\Phi}$ using the above architecture.
\comment{[CLOSED]
	Youssef: Any motivation behind why SIREN and not any other implicit
	representation. \\
	Patrick: How about the paragraph below?
}

For application of photon propagation, an optical visibility map $\vb{\Psi} :
\mathbb{R}^3 \mapsto \mathbb{R}^{N_d}$ is modeled using SIREN.
The function $\vb{\Psi}$ maps a coordinate $\vb{x} \in \mathbb{R}^3$ within the
detection volume to the visibility of $N_d$ optical detector for the simulated
ICARUS detector). There are three hyper-parameters for the SIREN model: number
of hidden layers ($n_L$) with sinusoidal activation, number of hidden features
($n_F$) and a frequency factor $\omega$.  In the models used for this work, all
hidden layers have the same number of hidden features: $M_i = N_i = n_F$. 

\comment{[FIXED] 
  Daniel: is there a reason to use the term "feature" rather than "neurons"
  here?  To me "feature" is typically used to describe the size of the input,
  not the model. I found this confusing in Fig.  10 \\
  Patrick: I followed the SIREN paper and used the term "hidden features".
} 

The frequency factor $\omega$ is introduced by factoring the weight matrix
$\vb{W} \to \omega \vb{W}$ in order to increase the spatial frequency of the
layers. This allows for a better match to the frequency spectrum of the
signal and accelerates the training of SIREN in all hidden
layers~\cite{sitzmann2020siren}, as discussed in Section~\ref{sec:hyperpars}.

As the parameterization $\vb{\Psi}$ is defined on the continuous domain of
$\vb{x}$, it is not limited by a voxel grid resolution, allowing for the
modeling of finer details than the standard lookup table photon library. It is
also more memory efficient than such a discrete lookup table.  The gradients
and higher order derivatives with respect to the input coordinates can be
computed in closed form instead of using the finite difference method. With the
well-behaved derivatives, SIREN offers additional applications such as solving
inverse problems. Further, as a neural network, SIREN is able to be optimized
via gradient descent.

\comment{[FIXED]
  Sean: any neural network is differentiable, e.g. with respect to parameters
  implies training. Why is SIREN different, and why is this useful for us? \\
  Daniel +1: could simply delete. \\
  Patrick: Deleted this sentence.
}
\outline{Model architecture and hyper-parameters}

\section{Optimization}
\subsection{Voxel-wise Loss}
\label{sec:voxel_wise_loss} 
\outline{How to train SIREN w/ photon lib.}
\comment{[FIXED]
  Sean: Don't forget comment from Daniel(?) about why we don't just train on
  these metrics. \\
  Daniel: and also consider moving the metrics to the front of
  this section, and then explain the need to change to v tilde for training. \\
  Patrick: Paragraphs rearranged.
}

\begin{figure}
  \centering
  \includegraphics[width=\linewidth]{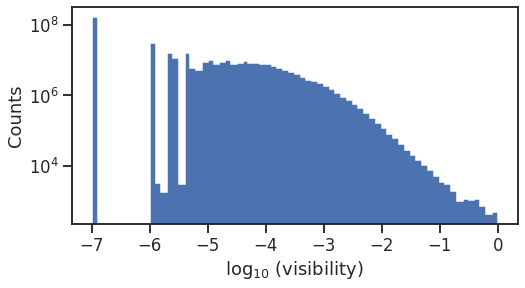}
  \caption{Distribution of the visibility of the ICARUS photon library. Data points 
    with zero visibility are grouped into the first bin.}
  \label{fig:icarus_vis}
\end{figure}

Let $\mathcal{D} = \{\vb{x}_i\} \subset \mathbb{R}^3$ be a set of voxel
coordinates and $v_{ij}$ be the visibility of position $\vb{x}_i$ at the $j$-th
photon detector. The values of $v_{ij}$ are obtained from the optical detector
simulation. The objective is to find a parameterization of the optical visibility
$\vb{\Psi}(\vb{x})$ that minimizes the absolute bias ($b_{\mathrm{abs}})$ and
relative bias ($b_\mathrm{rel}$), defined as
\begin{eqnarray}
  b_\mathrm{abs} &=& \left< | \vb{\Psi}_{j}(\vb{x}_i) - v_{ij} | \right>, \\
  b_\mathrm{rel} &=& \left<
    2 \left|
    \frac{\vb{\Psi}_{j}(\vb{x}_i) - v_{ij}}{\vb{\Psi}_{j}(\vb{x}_i) + v_{ij}}
    \right| \right>,
\end{eqnarray}
where $\left<\cdot\right>$ denotes the arithmetic mean. The calculation of
$b_\mathrm{abs}$ includes all data points and photomultiplier tubes (PMTs) in
the sample, while $b_\mathrm{rel}$ has a threshold of $v_{ij} > 4.5 \times
10^{-5}$. Because of the uneven distribution of $v_{ij}$ in the photon library
(Fig.~\ref{fig:icarus_vis}), $b_\mathrm{abs}$ is statistically weighted toward
the low-to-zero visibility region. In contrast, $b_\mathrm{rel}$ gives a better
measure of the accuracy of reproducing the bright detector regions with SIREN.

To account for the high dynamic range of the visibility, which typically spans
over several order of magnitude (Fig.~\ref{fig:icarus_vis}), $v_{ij}$ is
transformed into a  logarithm scale $\tilde{v}_{ij} = \log_{10}(v_{ij} +
\epsilon)$.  A small constant $\epsilon = 10^{-5}$ is added to avoid values of
infinity at $v_{ij} = 0$.  For training of the SIREN, the input coordinates,
$\vb{x}_i$, are normalized to lie within $[-1,1]$, and the logarithm of
visibility is normalized to $[0,1]$.

\comment{[FIXED] 
  Yifan: under equation(2), the definition of v ij is repeated from the previous text.
  It would be nice to give a sense of the scale of v ij, as you talk about low
  and high visibility and some threshold of it.
}

The training of SIREN minimizes the squared error:
\begin{equation} 
  \mathcal{L}_2 = \sum_{\vb{x}_i \in \mathcal{D}} \sum_{j=1}^{N_d} 
  w_{ij} [\tilde{\Psi}_j(\vb{x}_i) - \tilde{v}_{ij}] ^2, 
\end{equation}
where $w_{ij}$ is an optional precompuated weight, and
$\vb{\tilde{\Psi}}$ is a parameterizaion of the visibility after logarithmic
transformation as discussed above. The training sample $\mathcal{D}$ may be
obtained from a photon library lookup table.

\subsection{Track-wise Loss} 
\label{sec:track_wise_loss} 
\outline{How to train w/ tracks} 

The minimization of the voxel-wise loss $\mathcal{L}_2$ allows for a direct
representation of the photon library via sampling of the visibility at
different coordinates. However, the visibility of an individual voxel is not
available in real data because there is no point-like calibration source in the
LArTPC detector.  Therefore, an alternative approach is required for working
with data, where the only information is the readout of optical detectors from
particle tracks

\comment{[FIXED]
  Sean: I think the distinction between use cases of voxel-wise vs track-wise
  loss needs to be more clear -- maybe something like ``here's a direct
  parametrization of the current photon library (voxel), but for calibration to
  data we need an approach that works for data, here's how we would do that''.
  Daniel: +1 \\
  Patrick: Text modified.
}
The basic principle of the track-wise loss is to associate the flashes from the
optical detector to the charge readout of tracks in the LArTPC.  The number of
scintillation photons corresponding to an energy deposition in LAr is typically
$\mathcal{O}(10^4 - 10^5)$ photons/MeV depending on the magnitude of the drift
field~\cite{cennini1995}.  The expected number of photoelectrons (PEs)
$\lambda_j$ detected by the $j$-th PMT is given by
\begin{equation}
  \lambda_j = \sum_{t \in \mathcal{T}} 
    Y_j \times \Delta E_t \times \vb{\Psi}_j(\vb{x}_t),
  \label{eq:flash_pred}
\end{equation}
where $\mathcal{T}$ is the collection of charge voxels occupied by an image of
track(s), $Y_j$ is the PMT light yield, $\Delta E_t$ is the amount of the
energy deposited in LAr at voxel $t$, and $\Psi_j(\vb{x}_t)$ is the optical
visibility at the voxel coordinates $\vb{x}_t$. The PMT light yield ($Y_j$) is
a drift field dependent factor, which consists of the scintillation light
yield, the PMT collection efficiency and the PMT gain for the conversion to
photoelectrons. The typical value of $Y_j$ is $\mathcal{O}(1)$ PEs/MeV
at nominal drift field of 500 V/cm~\cite{sorel2014}.

A track-wise loss function is defined as the Poisson likelihood:
\begin{equation}
  \mathcal{L}_\mathrm{track} = \prod_{j=1}^{N_d}
    \mathrm{Pois}(n_j | \lambda_j),
\end{equation}
where $\mathrm{Pois}(n_j | \lambda_j) = {\lambda_j}^{n_j} \exp(-\lambda_j) /
n_j\!$ is the Poisson distribution of the observed number of photoelectrons in
the $j$-th PMT ($n_j$) given the expectation $\lambda_j$. The SIREN model
trained on simulatin using voxel-wise loss can be calibrated using track data
by minimizing the negative log-likelihood $-\ln\mathcal{L}_\mathrm{track}$ as
discussed in Sec.~\ref{sec:app}.

\comment{[FIXED]
  Sean: Motivate this loss function?
  Patrick: Added a sentence and forward reference to the "Application" section.
}
\section{Results} 
\label{sec:results}
\subsection{SIREN Representation}
\begin{figure}
  \centering
  \includegraphics[width=\linewidth]{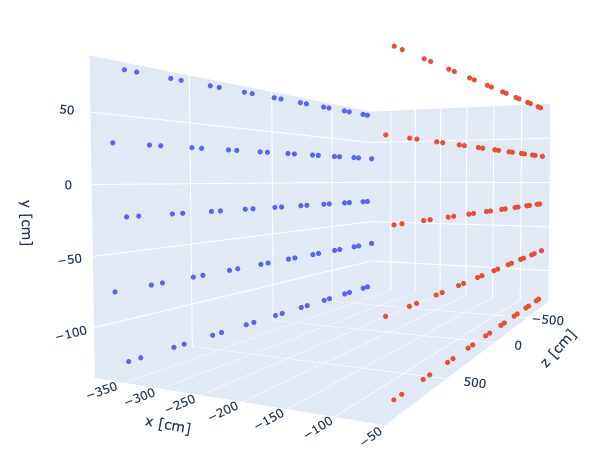}
  \caption{
    A module of the ICARUS detector. There are 180 PMTs mounted on the planes x
    = -59.36 mm (red) and -381.07 mm (blue), facing inwards to the detection
    volume.
  }
  \label{fig:icarus_detector}
\end{figure}

\begin{figure}
  \centering
  \includegraphics[width=\linewidth]{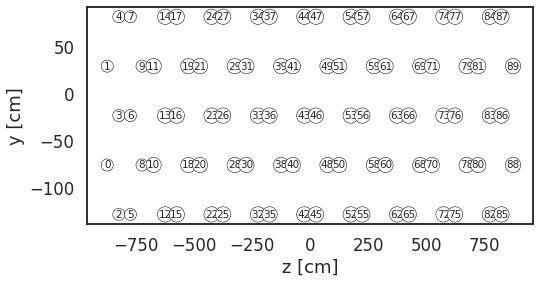}
  \caption{
    The locations of the first 90 PMTs at x = -381.07 mm of the ICARUS
    detector. The other 90 PMTs at x = -59.36 mm follow the same layout.
    Diameters of the PMTs are not drawn to scale.
  }
  \label{fig:icarus_pmt} 
\end{figure}

\outline{Description of ICARUS Photon Library} 
\comment{[FIXED] Yifan: 5mm, or 5cm? Patrick: 5cm}
A lookup table of the optical visibility is prepared for one module of the
ICARUS detector (Fig.~\ref{fig:icarus_detector}) with a total of 180
photomultiplier tubes (PMTs), 90 for each side of the drift volume
(Fig.~\ref{fig:icarus_pmt}).  The detection volume is divided into $74 \times
77 \times 394$ voxels with 5~cm in each dimension.
\comment{[FIXED] 
  Sean: 180 for each side, right? The text is ambiguous on that vs 180 total
} 
\comment{[FIXED] 
  Sean: These are very specific numbers, why this binning? \\
  Patrick: As pointed out in the intro., the binning is limited by the generation
  time. O(cm) is the smallest voxel size in practice. The bin numbers are simply
  the size of the detector volume by 5 cm.
} 
For each voxel, one million photons are generated isotropically and propagated
through the detection volume. The number of photons detected by each PMT is
recorded to estimate the optical visibility. The data from the resulting
visibility lookup table is then used to train the SIREN parameterization with
the voxel-wise loss ($\mathcal{L}_2$) described in
Section~\ref{sec:voxel_wise_loss}.
\comment{[FIXED]
	Youssef: why is Figure~\ref{fig:icarus_vis} here even though it was only
	referenced in the earlier sections? I think it should be Fig. 1. Also, why is
  this distribution discontinuous between -7 and -6?  
	Patrick: figures rearranged. The discontinuity is due to data points with vis=0.
	As mentioned in the caption, I put vis=0 into -7 bin.
}
\outline{Results for the baseline SIREN model}
\comment{[FIXED]
  Sean: Why is this the baseline (these hyperparameters, etc)? Might be
  worthwhile to forward reference the hyperparameter study. Also any reason for
  these batching choices?\\
  Patrick: Added a forward ref. \\
  The batch number is somewhat arbitrary. I picked an integral divisor of the total
  number of voxels. It also fits the GPU memory in the older 2080Ti.
}
A baseline SIREN model of $n_L=5$, $n_F=512$ and $\omega=30$ is trained on the
simulated ICARUS detector in a batch size of 60676 voxels, corresponding to a total 
of 37 batches per epoch. Other choices of hyperparameters are discussed in Sec.~\ref{sec:hyperpars}.
A weighting factor $w_{ij} \propto v_{ij}$ is included in the loss
function $\mathcal{L}_2$ to account for the statistical uncertainty on the
number of photons in the simulation. The training process takes about 30
minutes per 1000 epochs on a NVIDIA A100 GPU. 

\begin{figure}
  \centering
  \includegraphics[width=\linewidth]{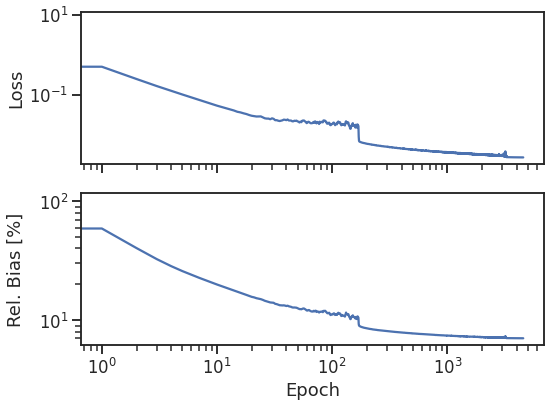}
  \caption{
    The loss (top) and relative bias (bottom) curves for training the baseline
    SIREN model with the ICARUS photon library. 
  }
  \label{fig:icarus_training}
\end{figure}

\begin{figure}
  \centering
  \includegraphics[width=\linewidth]{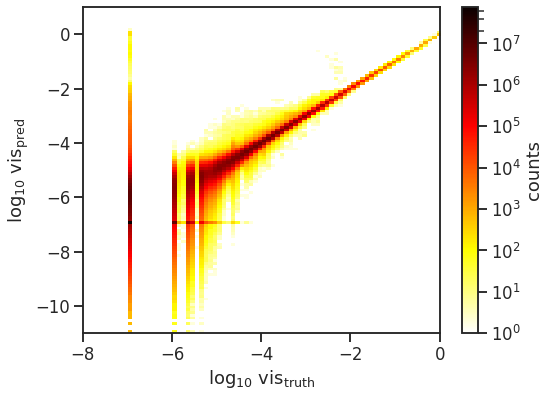}
  \caption{
    Predicted visibility from the baseline SIREN model v.s. the truth values from
    ICARUS photon library. Data points with zero visibility are put into the
    first non-empty bin. 
  }
  \label{fig:icarus_siren_scatter}
\end{figure}

\begin{figure}
  \centering
  \includegraphics[width=\linewidth]{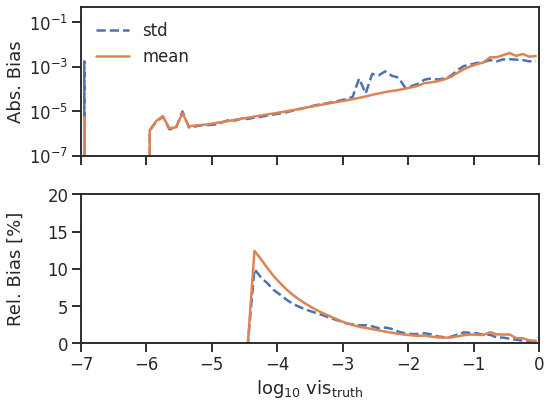}
  \caption{
    The mean (solid) and standard deviation (dashed) of the absolute (top) and
    relative (bottom) biases for the baseline SIREN model as a function of
    truth visibility from ICARUS photon library.  Data points with zero
    visibility are put into the first empty bin.
  } 
  \label{fig:icarus_bias}
\end{figure}

\begin{figure}
  \centering
  \includegraphics[width=\linewidth]{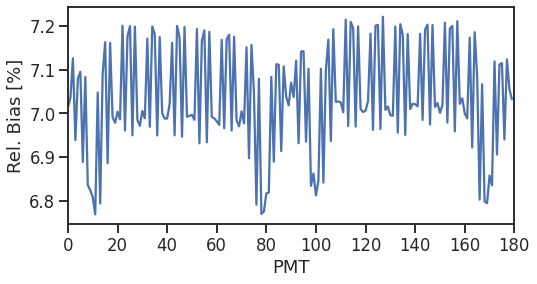}
  \caption{
    The relative bias of the baseline SIREN model, averaging over voxels, as a
    function of PMT indices. The variation of bias in PMTs is expected from the
    statistical unercainty of the simulation, as discussed in
    Section~\ref{sec:stat}.
  }
  \label{fig:icarus_bias_pmt}
\end{figure}

\comment{[FIXED]
  Sean: I think the point gets a little lost here -- might be worth it to
  headline as ``this 7\% comes mostly from regions with low visibility. As
  shown in Figures xx, for bright regions...'' and then forward reference with
  ``This is understood to be due to low statistics of detected photons, as
  demonstrated in section xx''\\
  Patrick: Text changed as suggested.
} 
The SIREN model converges quickly to a relative bias of 10\% within 200 epochs
and plateaus at 7\% (Fig.~\ref{fig:icarus_training}). As shown in
Figures~\ref{fig:icarus_siren_scatter} and \ref{fig:icarus_bias}, this 7\%
comes mostly from regions with low visibility.  For the bright regions with
visibility greater than $10^{-3}$, the SIREN model reproduces the photon
library within 1\%. This is understood to be due to low statistics of detected
photons in the simulation, as demonstrated in Section~\ref{sec:stat}.
Figure~\ref{fig:icarus_bias_pmt} shows a small varation of relative bias at
$\pm 0.2\%$ depending on the PMT locations. 
\comment{[FIXED?] 
  Sean: do we have an explanation for the structure here?
  Patrick: I think it is due to the geometry of the PMT arrangements.
  Unfortunately I don't have a quantitative explanation.
} 

\comment{[FIXED]
  Sean: This statement about Figure~\ref{fig:siren_result_xslice} feels a
  little out of place as is. Maybe rephrase as ``qualitative results'' or
  something?\\
  Patrick: Sentence rephrased.
}
\comment{[FIXED]
  Sean: You should emphasize that it's the choice of SIREN that allows the
  faithful reproduction of gradients. Do we care about this faithful
  reproduction, and do you have a quantitative measure of how well we do?\\
  Patrick: Edit w/ emphasis on the differentiablity of SIREN. We don't have
  benchmarks on the gradients. But the performance on flash matching
  demostrates it is good enough for physics application.
} 

\begin{figure}
  \centering
  \includegraphics[width=\linewidth]{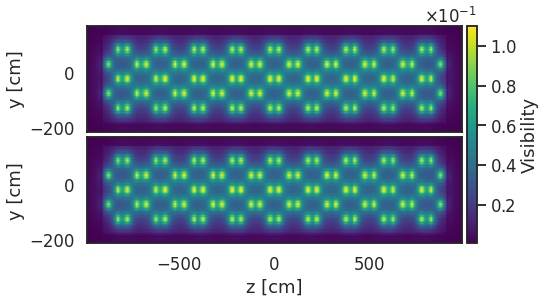}
  \caption{
    Sum of visibility for all PMTs at $x=-362.5~\mathrm{cm}$
    from the photon library (top) and the SRIEN model (bottom).
  }
  \label{fig:siren_vis}
\end{figure}

\begin{figure}
  \centering
  \includegraphics[width=\linewidth]{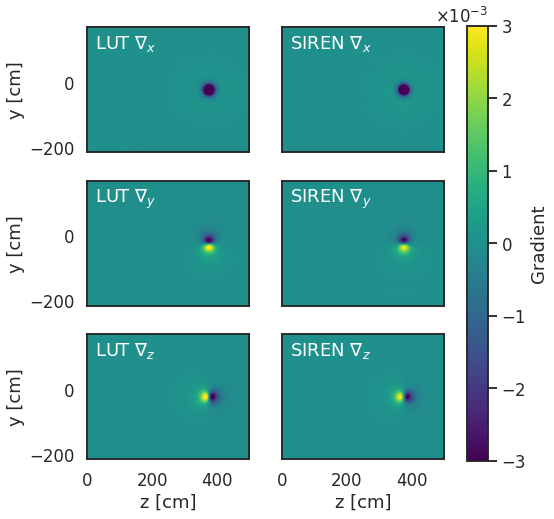}
  \caption{
    Gradients in $x$ (top), $y$ (middle) and $z$ (bottom) directions of PMT
    \#63 at $x=-362.5~\mathrm{cm}$ from the photon library (left) and
    the SIREN model (right). The gradients of the photon libraray are estimated
    using Sobel filter~\cite{sobel}, while the gradients of the SIREN model are
    computed using PyTorch’s automatic differentiation engine~\cite{pytorch}.
  } 
  \label{fig:siren_grad} 
\end{figure}

Figure~\ref{fig:siren_vis} shows that the SIREN model reproduces the
optical visibility map using a significantly smaller number of parameters
($\sim$1.4 million) than the lookup table approach ($\sim$404 million). Because
of the differentiablity with respect to the input coordinates of the SIREN
model, the gradient can be computed regardless of the grid
resolution~(Fig.\ref{fig:siren_grad}). The computation time and memory
footprint of the SIREN model make it amenable to the computing resources of
existing and future LArTPC experiments.  

\comment{[PENDING]
  Yifan: for Fig.~\ref{fig:icarus_siren_scatter}, would it be easy to add a
  grid on it to show it is highly diagonal.
}

\subsection{Statistical Uncertainty}
\label{sec:stat}
\begin{figure*}[btp]
  \centering
  \includegraphics[width=\textwidth]{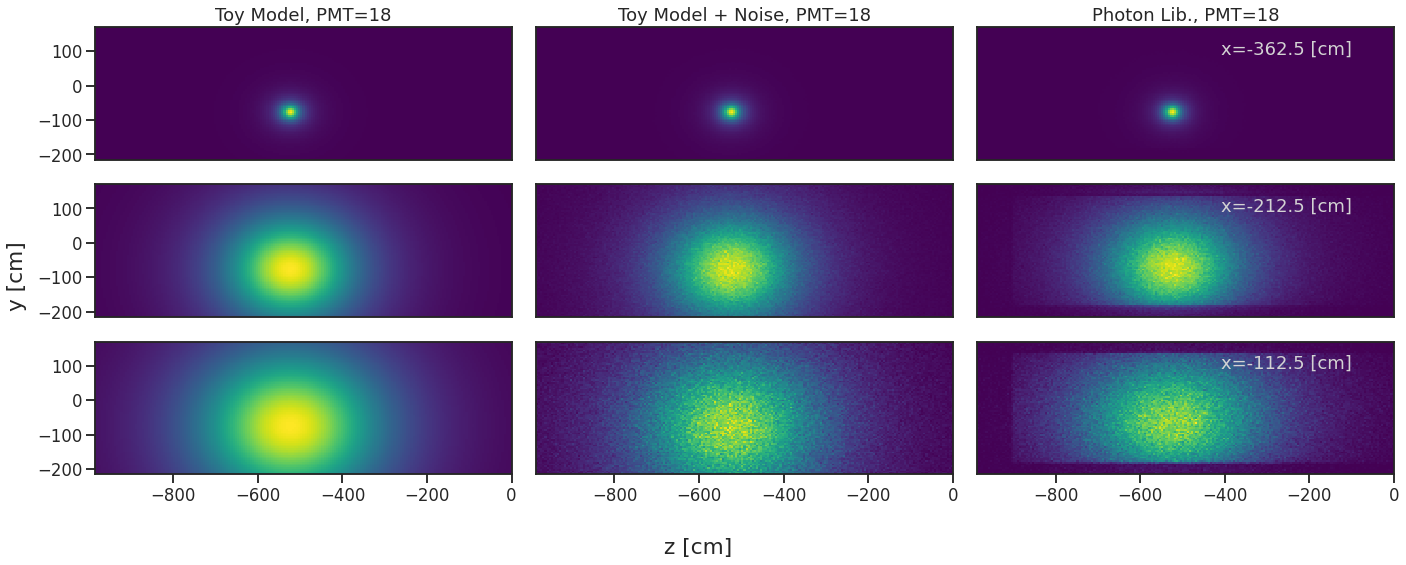}
  \caption{
    The visibility of PMT=18 at three different x positions for the toy model
    (left column), toy+noise sample (middle column) and the ICARUS photon
    library (right column).
  }
  \label{fig:toy_overview}
\end{figure*}

\begin{figure}[htbp]
  \centering
  \includegraphics[width=\linewidth]{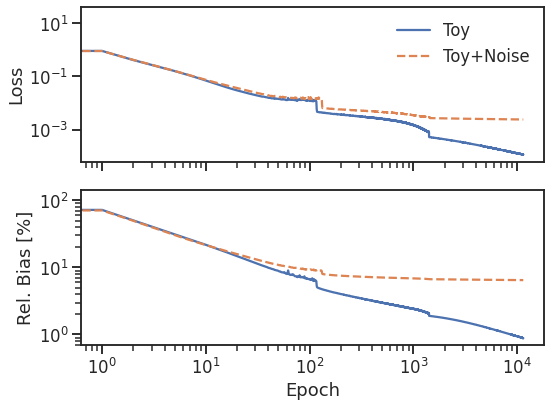}
  \caption{
    Training curves of the loss (top) and relative bias (bottom) for the toy
    model (solid) and the toy+noise sample (dashed).
  }
  \label{fig:toy_training}
\end{figure}

\begin{figure}[htbp]
  \centering
  \includegraphics[width=\linewidth]{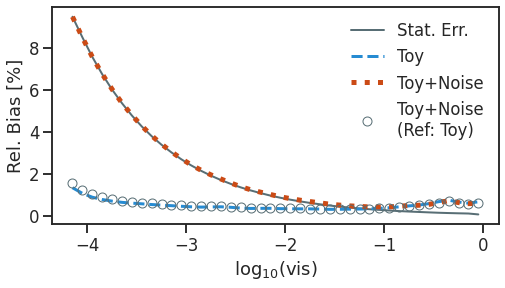}
  \caption{
    The relative bias for the analytical toy model (solid), toy+noise sample
    (red dotted) and expected statistical error (blue dashed) as a function of
    visibility. The data points (open circle) represents the SIREN model
    trained from the toy+noise sample with respect to the analytic toy model.
  }
  \label{fig:toy_bias}
\end{figure}

\outline{Photon library v.s. toy data (include statistical fluctuation in photon library)}
A finite number of simulated photons is used for the construction of the photon
library. There is therefore an inherent statistical limitation on the fidelity
of the photon library, in particular for the dark regions of the detector,
where very few photons are visible. As the photon library is used as the input
for training the SIREN model, this inherent statistical uncertainty translates
into an impact on the performance of the learned SIREN. To study this impact, a
{\em toy} photon library is generated analytically as
\begin{equation}
  \vb{\Psi}^\mathrm{toy}(\vb{x}) = \max(a e^{-k\vb{r(\vb{x})}}/\vb{r(\vb{x})}^2, 1),
\end{equation}
where $\vb{r}(\vb{x})$ is a vector of distances from the point $\vb{x}$ to the
PMTs. This toy model includes the two most important features of the photon
propagation, namely the inverse-square fall off away from the light source and the light
attenuation in the transportation media. The constants $a=55$ and $k=0.006$ are
chosen to loosely resemble the ICARUS photon library
(Fig.~\ref{fig:toy_overview}). A noisy visibility lookup table is generated by
\begin{equation} 
  v_{ij}^\mathrm{toy} = \max\{
  \mathrm{Pois}(\vb{\Psi}^\mathrm{toy}_j(\vb{x}_i) \times 10^6)/10^6, 1 \},
\end{equation}
where $\mathrm{Pois}(\cdot)$ denotes a Poisson random variable.  The toy+noise
sample mimics the global features of the ICARUS photon library, which has an
expected statistical uncertainty corresponding to $10^6$ generated photons per
voxel (Fig.~\ref{fig:toy_overview}).
\comment{[CLOSED]
  Yifan: above Eq.8, what about something like "To incorporate the PMT noises,
  an advanced toy lookup table is generated by...." \\
  Patrick: there is no PMT noise in the simulation. The "noisy" photon lib. only
  contains stat. err. on the number of detected photons.
}
The baseline SIREN training procedure is repeated for the toy model and
toy+noise photon library. Similar to Fig.~\ref{fig:icarus_training}, the toy
photon library with statistical uncertainty plateaus at 6.5\% in relative bias,
while the analytical toy model goes below 1\% and still shows a downward trend
after 10k epochs (Fig.~\ref{fig:toy_training}). As shown in
Fig.~\ref{fig:toy_bias}, the relative bias of the SIREN trained on the
toy+noise sample follows the expected statistical error when compared to the
toy+noise sample used for training, showing a similar trend as observed in
Fig.~\ref{fig:icarus_bias}. 

\comment{[FIXED]
  Sean: maybe slight rephrase to just say generally ``the bias scales with
  statistical error, demonstrating a similar trend as a function of visibility as
  that seen in Section xx''? \\
  Patrick: as suggested.
}

\begin{figure}[htbp]
  \centering
  \includegraphics[width=\linewidth]{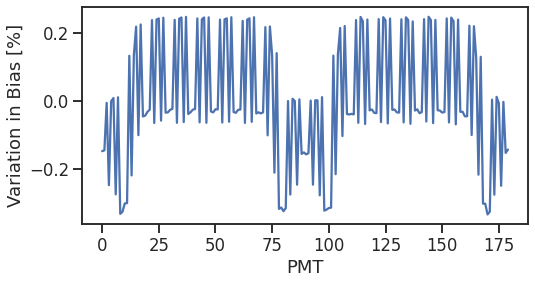}
  \caption{
    Expected variation in bias for different PMTs due to the statistical uncertainty from simulation. 
  }
  \label{fig:exp_bias}
\end{figure}

On the other hand, the relative bias of the SIREN trained on the toy sample has
very little dependence on the visibility at $\sim$1\%
(Fig.~\ref{fig:toy_bias}). By comparing the SIREN parameterization of the
toy+noise sample to the toy model, the relative bias is similar to the SIREN
model trained using toy model without statistical uncertainty. It indicates
that the SIREN model is able to remove the statistical fluctuation from the
photon library and learn the underlying distribution. It is therefore a more
robust model of the visibility than the voxel representation, in addition to
the other benefits as described above. The expected variation in bias due to
statistical uncertainty from simulation is shown in Fig.~\ref{fig:exp_bias}.
The performance of SIREN modeling of the ICARUS's visibility LUT
(Fig.~\ref{fig:icarus_bias_pmt}) is dominated by the statistical fluctuation
inheriting from the simulation.

\comment{[CLOSED]
  Sean: wait, I'm confused -- the fact that SIREN trained on Toy+Noise doesn't
  scale with uncertainty is \emph{different} from what we see above, right?
  This seems to say that SIREN is not impacted by statistical uncertainty, but
  we started off by saying that it is
}
\comment{[CLOSED]
  Yifan: the legend of Fig 13 is confusing. Actually I don't think I understand 
  the plot.
}
\comment{[FIXED]
  Figure~\ref{fig:toy_bias} updated. Please check the above paragraphs and make
  sure the follow messages are delivered: \\
  1) majority of the 7\% bias observed in the ICARUS SIREN model is due to stat. 
  uncertainty,\\
  2) SIREN is a better approach than LUT because the ability to recover the 
  underlying distribution. \\
	Sean: I think this is better, but maybe could be made clearer by being a bit
  more explicit, e.g. ``the relative bias of the toy+noise sample...''
  $\rightarrow$ ``the relative bias of the SIREN trained on the toy+noise sample
  follows the expected statistical error when compared to the toy+noise sample
  used for training.'', and same type of thing in other places\\
	Patrick: As suggested.
}

\subsection{Hyperparameters}
\label{sec:hyperpars}
\outline{Study of hyperparameters} 

\begin{figure}
  \centering
  \includegraphics[width=\linewidth]{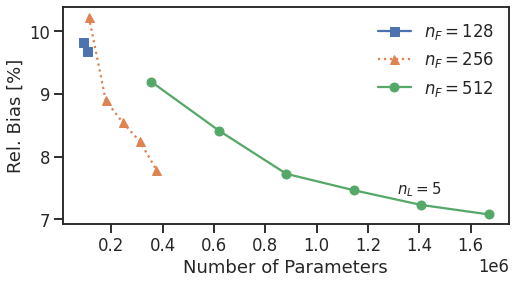}
  \caption{The best relative bias v.s. number of parameters for different SIREN 
  hyperparameters of $n_F=128,~n_L=\{4,5\}$ (squares), $n_F=256,~n_L=\{1,2,3,4,5\}$ 
  (triangles), and $n_F=512,~n_L=\{1,2,3,4,5\}$ (circles).}
  \label{fig:hyperpars}
\end{figure}

Different choices of hyperparameters are explored and compared to the baseline
setting ($n_L=5, n_F=512, \omega=30$). Since the construction of SIREN from
photon library is an overfitting problem, the accuracy will alwyas increase
with the depth of the network and the number of hidden features. As shown in
Fig.~\ref{fig:hyperpars}, the chosen baseline setting gives a bias that is
consistent with the expected statistical uncertainty (Fig.~\ref{fig:toy_bias}).
Adding one more layer ($n_L=6$) only gives a marginal improvement.

\begin{figure}
  \centering
  \includegraphics[width=\linewidth]{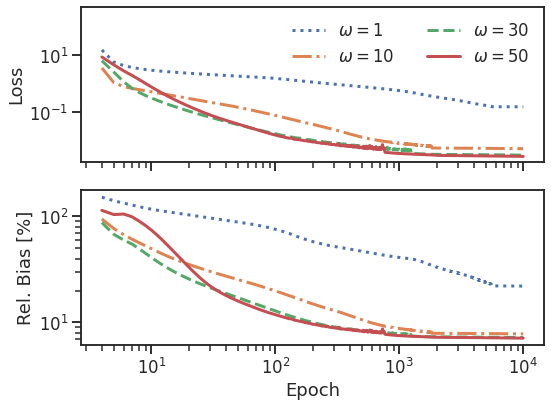}
  \caption{A study of hyperparameter $\omega$ for $n_F=512$ and $n_L=5$.}
  \label{fig:oemga}
\end{figure}

As shown in Fig.~\ref{fig:oemga}, the frequency factor $\omega$ has significant
impact on the SIREN training time~\cite{sitzmann2020siren}. Without this
multiplying factor ($\omega=1$), the convergence is much slower than with
$\omega>1$. The choice of $\omega=30$ is optimal for the training. There is no
further improvement with larger values. 

\comment{[FIXED]
  The more I work on this section, the more struggle I have. I couldn't find a
  compelling argument why we chose the baseline model. It just happeded the the
  same setting from the SIREN paper also seems to work for our application as
  well. There are two points behind the scence: \\
  1) SIREN is a over-fitting problem. There is no optimal hyperparameters. The 
  more number of parameters, the better the fit. \\
  2) If we keep increasing the number of parameters, I'm afraid the SIREN also
  starts learning/memorizing the pattern of statistical noise as well, which 
  does not contain any physics. Since we don't know the underlaying distribution,
  it is difficult to find a balance setting. \\
  There is a new plot in Fig.~\ref{fig:hyperpars} as suggested by Daniel. \\
  Please help to refine this section.\\
	Sean: I think it's fine to just say ``we used these'' for the parameters you
  don't study, and what you have for number of layers and features seems okay to
  me as is
	Patrick: subsection rewritten.
}
\comment{[PENDING]
  Patrick: This part should be updated to be consistent to the previous
  section. i.e. the target bias should 7\% instead of 8.2\%. The reason
  we picked 8.2\% because it was the best result from the previous study.
}
\comment{[CLOSED]
  Daniel: I don't think you need this figure -- if I understand, this just
  shows how many parameters there are vs. the size of the hidden layer. It would
  be more useful if e.g. you plotted the number of parameters vs. the training
  time or the bias.  Or maybe you don't need it at all.  \\
  Patrick: Replaced with Fig.~\ref{fig:hyperpars}. But I'm not what to conclude
  (see above).
}

\section{Applications}
\label{sec:app}
\subsection{Flash Matching} 
\comment{[FIXED] 
  Yifan: A suggestion, "It is an intrinsic challenge to use only charge
  information from LArTPCs to determine the position of particles along the
  drift direction of the detector. Additional optical detectors (e.g PMTs) can
  detect scintillation light from particles in LArTPCs and provide timestamps
  of nanosecond precision, which can be used to accurately project the charge
  signals along the drift direction." \\
  Sean: I like this, just made a small language edit \\
  Patrick: Text replaced.
}
It is an intrinsic challenge to use only charge information from LArTPCs to
determine the position of particles along the drift direction of the detector.
Additional optical detectors (e.g PMTs) can detect scintillation light from
particles in LArTPCs and provide timestamps of nanosecond precision, which can
be used to accurately project the charge signals along the drift direction.
\comment{[FIXED] Yifan: L333, 
  "The corresponding PMT readout given the charge signal is modeled using Eq. 5."
}
The flash matching algorithm assumes a matched image pair from the charge
readout and the PMT system. First, an offset $(x,y,z) \mapsto (x+x_0,y,z)$ is
applied along the drift direction for all of the voxels in the charge readout.
Then the corresponding PMT readout given the charge signal is modeled using
Eq.~\ref{eq:flash_pred}.  The goal is to minimize
$-\ln\mathcal{L}_\mathrm{track}$ for the offset $x_0$ using gradient descent
, while keeping all other parameters fixed.

\comment{[FIXED] Sean: What do you actually minimize? A NLL?}
\comment{[FIXED]
  Yifan: "The flash matching algorithm is implemented three ways to
  benchmark the performance of the SIREN model."?\\ 

  Sean: Agreed that current
  phrasing is strange. Another proposal is start with description of SIREN
  version and present the other two as comparison baselines\\
  Patrick: Paragraph rewritten. Yes, all three algorithms should perform the
  same, as the test sample is generated from the same ICARUS photon library.
  I'd prefer stick with the flow (history) on the development of these algos.,
  i.e. \\ 
  1) C++/LUT, 2) PyTorch/LUT, 3) PyTorch/SIREN
}
\comment{[CLOSE?]
  Please check the updated text. The selling points are the scalability of SIREN, 
  and the continous parameterization and gradient calculation. \\
	Sean: Maybe I'm being slow, but I still don't fully understand the
  differences here -- what exactly do you mean by ``calculated from the grid
  values of the photon library''? \\
	Patrick: To estimate the gradient of the LUT, you need to the grid data points
	$\{i-1,i,i+1\}\times\{j-1,j,j+1\}$. For SIREN, it is a continous function, and
	the derivatives w.r.t. to input coordinates are well-defined.
}

\comment{[FIXED]
  Yifan: L346 "This approach depends heavily on the numerical evaluation of the
  derivative of Ltrack." Isn't it true for SIREN as well, just that the
  information has been transformed? \\

  Patrick: What I meant was the calculation of gradient based on the grids of
  the photon lib. SIREN is differentiable by construction, i.e. d sin() = cos().
  Changed some wordings.
}

Traditionally the flash matching algorithm, as implemented by OpT0Finder, uses
a C++ optimization library called MINUIT~\cite{james1998} . As an intermediate
benchmark, we reimplemented the OpT0Finder algorithm in Python, and the MINUIT
library is replaced by PyTorch's automatic differentiation engine. Both the
OpT0Finder and PyTorch methods use the lookup table as input, where the
gradient is calculated from the grid values of the photon library using finite
differences. The SIREN model is incorporated into the PyTorch implementation to
replace the photon library, where the gradient of the visibilty is given by the
derivative of the SIREN model.

\begin{figure}[hbtp!]
  \centering
  \includegraphics[width=\linewidth]{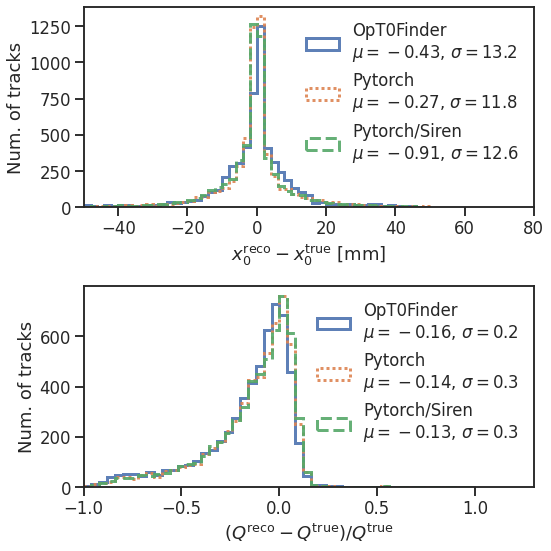}
  \caption{
    Benchmarks of the flash matching algorithm implemented in OpT0Finder,
    Pytorch and PyTorch+SIREN (see text for descriptions) in terms of
    reconstructed position (top) and reconstructed number of photoelectrons
    in PMTs (bottom).
  }
  \label{fig:flash_matching_benchmark}
\end{figure}

A test sample of 10000 tracks in random locations and the corresponding photon
detector output is generated according to Eq.~\ref{eq:flash_pred} using the
ICARUS photon library. The three flash matching algorithms are benchmarked with
the track samples. As shown in Fig.~\ref{fig:flash_matching_benchmark}, all
three implementations give comparable results for the reconstruction of the
absolute positions of the tracks and the number of photoelectrons of the PMTs.
Though the performance is similar for all methods, the SIREN model is much more
scalable than the lookup table approach, as it does not require grid-based
calculation of the gradient.

\comment{[FIXED]
  Sean: How is the differentiability of the SIREN model used here? As described, 
  you can do the same fitting with a lookup table, so not sure this is relevant \\
  Patrick: Text updated. 
}

\subsection{Data-driven Calibration}
\begin{figure}[http!]
  \centering
  \includegraphics[width=\linewidth]{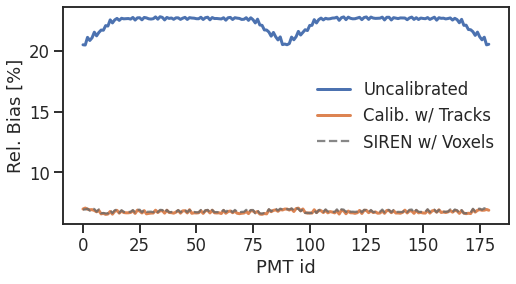}
  \caption{
    The relative bias of the modified photon library calculated using an
    uncalibrated SIREN model (blue).  The calibrated SIREN model (orange) using
    10000 tracks is consistent with re-training SIREN model with all the voxels
    from the modified photon library.
  }
  \label{fig:calib}
\end{figure}
\comment{[FIXED]
  Sean: so if I'm understanding, you're training the SIREN with voxel loss,
  then tuning with track loss (update: yes, I guess this is said later on)?
  Might be good to ref those sections and then rephrase, e.g., ``The SIREN
  models presented in Section xx are trained using a voxel-based loss to
  parametrize a fixed photon library lookup table. As this photon library is
  generated via detector simulation, a calibration procedure is required to
  correct for differences with respect to measured data. Because the learned
  SIREN model is fully differentiable, it may be automatically tuned to the
  data via a track-wise loss, as described in Section yy, a notable advantage
  of our approach.'' \\

  Patrick: Text replaced.
}
The SIREN models presented in Section~\ref{sec:voxel_wise_loss} are trained
using a voxel-based loss to parametrize a fixed photon library lookup table. As
this photon library is generated via detector simulation, a calibration
procedure is required to correct for differences with respect to measured data.
Because the learned SIREN model is fully differentiable, it may be
automatically tuned to the data via a track-wise loss, as described in
Section~\ref{sec:track_wise_loss}, a notable advantage of the SIREN approach.

The calibration process follows a similar optimization procedure as the flash
matching algorithm, where the goal again is to minimize
$-\ln\mathcal{L}_\mathrm{track}$.  However in this case, x0 is treated as
fixed, and the SIREN parameters are optimized The assumption of known $x_0$ can
be satisfied by selecting tracks crossing the physical boundaries of the
detector.

A modified photon library is generated by multiplying $\exp(-k r(\vb{x}))$ to
the original ICARUS photon library, where $r$ is the distance between position
$\vb{x}$ and a PMT and $k = 10^{-3}~\mathrm{cm}^{-1}$. The modified photon
library has a maximum reduction of 85\% in visibility and causes an average bias
of 22.3\% compared the SIREN model trained with the original ICARUS photon
library (Fig.~\ref{fig:calib}).

For the demonstration of the calibration process, a dataset of 10000 images of
the photon detector are generated from simulating single track per image and
the modified photon library according to Eq.~\ref{eq:flash_pred}. The track
sample covers about one-third of the voxel space of the ICARUS detector volume.
The SIREN model trained from the original ICARUS photon library is further
fine-tuned with the track sample. As shown in Fig.~\ref{fig:calib}, the
calibrated SIREN model performs the same as if it was trained from scratch
using the voxel-wise loss on the modified photon library.

\comment{[PENDING]
  Daniel: As we discussed in the meeting, it would be good to have an explicit
  comparison of the SIREN to the look-up table. I don't recall where we ended
  up in this discussion (maybe it's recorded?) but off the top of my head you
  could compare number of parameters (or equivalently storage size) and
  accuracy. I think I understand that accuracy is limited by the noise in the
  training dataset, so presumably is the same for both methods, but maybe the
  SIREN can learn to smooth and is even better than the lookup table?  If not
  limited by noise, you could also compare the error in the SIREN to the lookup
  table error due to pixelation (i.e. difference between neighboring pixels). \\
  ---\\
  Patrick: Figure~\ref{fig:hyperpars} gives the accuracy v.s. nuumber of 
  parameters. Also the updated version of Fig.~\ref{fig:toy_bias} demonstrates 
  that SIREN can learn to smooth the stat. err, and hence better than the lookup 
  table.\\
  Regarding the voxelization error, I don't how to proceed. Maybe training SIREN 
  from a subset of the lookup table and compare to the whole photon library? 
  Any further suggestiong?
}

\comment{[CLOSED?]
  Daniel: Related to above, In addition to number of parameters and accuracy,
  is it also useful to compare creation time, i.e. initial calculation for
  lookup table vs. training for SIREN? How does the size of the training set
  for the SIREN compare to the baseline lookup table? I guess in this example
  the model is trained on the equivalent lookup table, so the SIREN savings
  come purely during inference? Or there's the presumption that you will
  interpolate to higher resolution? \\
  --- \\
  Patrick: A realistic workflow for an experiement is to train a SIREN model
  with a lookup table (yes, the train sample = lookup table), and then
  calibrate (tune) the SIREN model with track data. This is one of the main
  advantage of SIREN over the lookup table. The timing saving part is that we
  don't need to regenerate the photon library after calibration.  There is
  interpolataion during inference, e.g. using the flash matching algorithm to
  locate the absolute position $x_0$. 
}
\section{Conclusion}
\comment{[CLOSED] Please help shaping up the conclusion.
Sean: Made a few edits -- just trying to emphasize the problems that we solve a bit more}

We propose a new approach to parametrize LArTPC optical visibility using SIREN,
a neural implicit representation with periodic activation functions. We
demonstrate that this approach is able to  reproduce the photon acceptance map
with high accuracy, and further show that this approach is less sensitive to
simulation statistics than commonly used lookup table methods. The number of
parameters in our SIREN model is orders of magnitude smaller than the number of
voxels in such lookup tables, making SIREN much more scalable to larger
detectors.  Furthermore, the SIREN model is easily tunable via automatic
differentiation, and has well behaved derivatives due to its periodic
activation functions. We demonstrate the potential of using these qualities to
optimize SIREN directly on real data, an application which is infeasible with
traditional approaches, mitigating the concern of data v.s. simulation
discrepancies. We further present an application for data reconstruction where
SIREN is used to form a likelihood function for photon statistics. 

In summary, our method offers a pathway towards improving the physics quality
of LArTPC simulation, while at the same time addressing issues of scalability
which are essential problems for the future of LArTPC experiments. We note that
this method is just one example of the power of differentiable surrogates in
physics simulation, and we hope that it prompts ideas for the use of such
methods in other areas of physics.

\section{Acknowledgement}
The authors wish to thank the ICARUS collaboration for providing access to the
photon detector visibility lookup table upon which the results presented in
Section~\ref{sec:results} are based.  This work is supported by the U.S.
Department of Energy, Office of Science, Office of High Energy Physics, and
Early Career Research Program under Contract DE-AC02-76SF00515.

\bibliography{references}{}

\begin{thebibliography}{14}%
\makeatletter
\providecommand \@ifxundefined [1]{%
 \@ifx{#1\undefined}
}%
\providecommand \@ifnum [1]{%
 \ifnum #1\expandafter \@firstoftwo
 \else \expandafter \@secondoftwo
 \fi
}%
\providecommand \@ifx [1]{%
 \ifx #1\expandafter \@firstoftwo
 \else \expandafter \@secondoftwo
 \fi
}%
\providecommand \natexlab [1]{#1}%
\providecommand \enquote  [1]{``#1''}%
\providecommand \bibnamefont  [1]{#1}%
\providecommand \bibfnamefont [1]{#1}%
\providecommand \citenamefont [1]{#1}%
\providecommand \href@noop [0]{\@secondoftwo}%
\providecommand \href [0]{\begingroup \@sanitize@url \@href}%
\providecommand \@href[1]{\@@startlink{#1}\@@href}%
\providecommand \@@href[1]{\endgroup#1\@@endlink}%
\providecommand \@sanitize@url [0]{\catcode `\\12\catcode `\$12\catcode
  `\&12\catcode `\#12\catcode `\^12\catcode `\_12\catcode `\%12\relax}%
\providecommand \@@startlink[1]{}%
\providecommand \@@endlink[0]{}%
\providecommand \url  [0]{\begingroup\@sanitize@url \@url }%
\providecommand \@url [1]{\endgroup\@href {#1}{\urlprefix }}%
\providecommand \urlprefix  [0]{URL }%
\providecommand \Eprint [0]{\href }%
\providecommand \doibase [0]{http://dx.doi.org/}%
\providecommand \selectlanguage [0]{\@gobble}%
\providecommand \bibinfo  [0]{\@secondoftwo}%
\providecommand \bibfield  [0]{\@secondoftwo}%
\providecommand \translation [1]{[#1]}%
\providecommand \BibitemOpen [0]{}%
\providecommand \bibitemStop [0]{}%
\providecommand \bibitemNoStop [0]{.\EOS\space}%
\providecommand \EOS [0]{\spacefactor3000\relax}%
\providecommand \BibitemShut  [1]{\csname bibitem#1\endcsname}%
\let\auto@bib@innerbib\@empty
\bibitem [{\citenamefont {Acciarri}\ \emph {et~al.}()\citenamefont {Acciarri}
  \emph {et~al.}}]{acciarri2015proposal}%
  \BibitemOpen
  \bibfield  {author} {\bibinfo {author} {\bibfnamefont {R.}~\bibnamefont
  {Acciarri}} \emph {et~al.},\ }\href@noop {} {\ }\Eprint
  {http://arxiv.org/abs/1503.01520} {arXiv:1503.01520} \BibitemShut {NoStop}%
\bibitem [{\citenamefont {Abud}\ \emph {et~al.}()\citenamefont {Abud} \emph
  {et~al.}}]{abud2021deep}%
  \BibitemOpen
  \bibfield  {author} {\bibinfo {author} {\bibfnamefont {A.}~\bibnamefont
  {Abud}} \emph {et~al.} (\bibinfo {collaboration} {DUNE}),\ }\href@noop {} {\
  }\Eprint {http://arxiv.org/abs/2103.13910} {arXiv:2103.13910} \BibitemShut
  {NoStop}%
\bibitem [{\citenamefont {Acciarri}\ \emph {et~al.}(2017)\citenamefont
  {Acciarri} \emph {et~al.}}]{acciarri2017design}%
  \BibitemOpen
  \bibfield  {author} {\bibinfo {author} {\bibfnamefont {R.}~\bibnamefont
  {Acciarri}} \emph {et~al.} (\bibinfo {collaboration} {MicroBooNE}),\ }\href
  {https://doi.org/10.1088/1748-0221/12/02/p02017} {\bibfield  {journal}
  {\bibinfo  {journal} {JINST}\ }\textbf {\bibinfo {volume} {12}},\ \bibinfo
  {pages} {P02017} (\bibinfo {year} {2017})}\BibitemShut {NoStop}%
\bibitem [{\citenamefont {Dwyer}\ \emph {et~al.}(2018)\citenamefont {Dwyer},
  \citenamefont {Garcia-Sciveres}, \citenamefont {Gnani}, \citenamefont
  {Grace}, \citenamefont {Kohn}, \citenamefont {Kramer}, \citenamefont
  {Krieger}, \citenamefont {Lin}, \citenamefont {Luk}, \citenamefont {Madigan},
  \citenamefont {Marshall}, \citenamefont {Steiner},\ and\ \citenamefont
  {Stezelberger}}]{dwyer2018larpix}%
  \BibitemOpen
  \bibfield  {author} {\bibinfo {author} {\bibfnamefont {D.}~\bibnamefont
  {Dwyer}}, \bibinfo {author} {\bibfnamefont {M.}~\bibnamefont
  {Garcia-Sciveres}}, \bibinfo {author} {\bibfnamefont {D.}~\bibnamefont
  {Gnani}}, \bibinfo {author} {\bibfnamefont {C.}~\bibnamefont {Grace}},
  \bibinfo {author} {\bibfnamefont {S.}~\bibnamefont {Kohn}}, \bibinfo {author}
  {\bibfnamefont {M.}~\bibnamefont {Kramer}}, \bibinfo {author} {\bibfnamefont
  {A.}~\bibnamefont {Krieger}}, \bibinfo {author} {\bibfnamefont
  {C.}~\bibnamefont {Lin}}, \bibinfo {author} {\bibfnamefont {K.}~\bibnamefont
  {Luk}}, \bibinfo {author} {\bibfnamefont {P.}~\bibnamefont {Madigan}},
  \bibinfo {author} {\bibfnamefont {C.}~\bibnamefont {Marshall}}, \bibinfo
  {author} {\bibfnamefont {H.}~\bibnamefont {Steiner}}, \ and\ \bibinfo
  {author} {\bibfnamefont {T.}~\bibnamefont {Stezelberger}},\ }\href
  {https://doi.org/10.1088/1748-0221/13/10/p10007} {\bibfield  {journal}
  {\bibinfo  {journal} {JINST}\ }\textbf {\bibinfo {volume} {13}},\ \bibinfo
  {pages} {P10007} (\bibinfo {year} {2018})}\BibitemShut {NoStop}%
\bibitem [{\citenamefont {Rubbia}\ \emph {et~al.}(2011)\citenamefont {Rubbia}
  \emph {et~al.}}]{rubbia2011icarus}%
  \BibitemOpen
  \bibfield  {author} {\bibinfo {author} {\bibfnamefont {C.}~\bibnamefont
  {Rubbia}} \emph {et~al.} (\bibinfo {collaboration} {ICARUS}),\ }\href
  {https://doi.org/10.1088/1748-0221/6/07/p07011} {\bibfield  {journal}
  {\bibinfo  {journal} {JINST}\ }\textbf {\bibinfo {volume} {6}},\ \bibinfo
  {pages} {P07011} (\bibinfo {year} {2011})}\BibitemShut {NoStop}%
\bibitem [{\citenamefont {Abi}\ \emph {et~al.}(2020)\citenamefont {Abi} \emph
  {et~al.}}]{abi2020tdr4}%
  \BibitemOpen
  \bibfield  {author} {\bibinfo {author} {\bibfnamefont {B.}~\bibnamefont
  {Abi}} \emph {et~al.} (\bibinfo {collaboration} {DUNE}),\ }\href
  {https://doi.org/10.1088/1748-0221/15/08/t08010} {\bibfield  {journal}
  {\bibinfo  {journal} {JINST}\ }\textbf {\bibinfo {volume} {15}},\ \bibinfo
  {pages} {T08010} (\bibinfo {year} {2020})}\BibitemShut {NoStop}%
\bibitem [{\citenamefont {Mildenhall}\ \emph {et~al.}()\citenamefont
  {Mildenhall}, \citenamefont {Srinivasan}, \citenamefont {Tancik},
  \citenamefont {Barron}, \citenamefont {Ramamoorthi},\ and\ \citenamefont
  {Ng}}]{mildenhall2020nerf}%
  \BibitemOpen
  \bibfield  {author} {\bibinfo {author} {\bibfnamefont {B.}~\bibnamefont
  {Mildenhall}}, \bibinfo {author} {\bibfnamefont {P.~P.}\ \bibnamefont
  {Srinivasan}}, \bibinfo {author} {\bibfnamefont {M.}~\bibnamefont {Tancik}},
  \bibinfo {author} {\bibfnamefont {J.~T.}\ \bibnamefont {Barron}}, \bibinfo
  {author} {\bibfnamefont {R.}~\bibnamefont {Ramamoorthi}}, \ and\ \bibinfo
  {author} {\bibfnamefont {R.}~\bibnamefont {Ng}},\ }\href@noop {} {\ }\Eprint
  {http://arxiv.org/abs/2003.08934} {arXiv:2003.08934} \BibitemShut {NoStop}%
\bibitem [{\citenamefont {Martel}\ \emph {et~al.}()\citenamefont {Martel},
  \citenamefont {Lindell}, \citenamefont {Lin}, \citenamefont {Chan},
  \citenamefont {Monteiro},\ and\ \citenamefont {Wetzstein}}]{martel2021acron}%
  \BibitemOpen
  \bibfield  {author} {\bibinfo {author} {\bibfnamefont {J.~N.~P.}\
  \bibnamefont {Martel}}, \bibinfo {author} {\bibfnamefont {D.~B.}\
  \bibnamefont {Lindell}}, \bibinfo {author} {\bibfnamefont {C.~Z.}\
  \bibnamefont {Lin}}, \bibinfo {author} {\bibfnamefont {E.~R.}\ \bibnamefont
  {Chan}}, \bibinfo {author} {\bibfnamefont {M.}~\bibnamefont {Monteiro}}, \
  and\ \bibinfo {author} {\bibfnamefont {G.}~\bibnamefont {Wetzstein}},\
  }\href@noop {} {\ }\Eprint {http://arxiv.org/abs/2105.02788}
  {arXiv:2105.02788} \BibitemShut {NoStop}%
\bibitem [{\citenamefont {Sitzmann}\ \emph {et~al.}()\citenamefont {Sitzmann},
  \citenamefont {Martel}, \citenamefont {Bergman}, \citenamefont {Lindell},\
  and\ \citenamefont {Wetzstein}}]{sitzmann2020siren}%
  \BibitemOpen
  \bibfield  {author} {\bibinfo {author} {\bibfnamefont {V.}~\bibnamefont
  {Sitzmann}}, \bibinfo {author} {\bibfnamefont {J.~N.~P.}\ \bibnamefont
  {Martel}}, \bibinfo {author} {\bibfnamefont {A.~W.}\ \bibnamefont {Bergman}},
  \bibinfo {author} {\bibfnamefont {D.~B.}\ \bibnamefont {Lindell}}, \ and\
  \bibinfo {author} {\bibfnamefont {G.}~\bibnamefont {Wetzstein}},\ }\href@noop
  {} {\ }\Eprint {http://arxiv.org/abs/2006.09661} {arXiv:2006.09661}
  \BibitemShut {NoStop}%
\bibitem [{\citenamefont {Cennini}\ \emph {et~al.}(1995)\citenamefont
  {Cennini}, \citenamefont {Revol}, \citenamefont {Rubbia}, \citenamefont
  {Tian}, \citenamefont {{Dzialo Giudice}}, \citenamefont {Li}, \citenamefont
  {Motto}, \citenamefont {Picchi}, \citenamefont {Boccaccio}, \citenamefont
  {Cavanna} \emph {et~al.}}]{cennini1995}%
  \BibitemOpen
  \bibfield  {author} {\bibinfo {author} {\bibfnamefont {P.}~\bibnamefont
  {Cennini}}, \bibinfo {author} {\bibfnamefont {J.}~\bibnamefont {Revol}},
  \bibinfo {author} {\bibfnamefont {C.}~\bibnamefont {Rubbia}}, \bibinfo
  {author} {\bibfnamefont {W.}~\bibnamefont {Tian}}, \bibinfo {author}
  {\bibfnamefont {D.}~\bibnamefont {{Dzialo Giudice}}}, \bibinfo {author}
  {\bibfnamefont {X.}~\bibnamefont {Li}}, \bibinfo {author} {\bibfnamefont
  {S.}~\bibnamefont {Motto}}, \bibinfo {author} {\bibfnamefont
  {P.}~\bibnamefont {Picchi}}, \bibinfo {author} {\bibfnamefont
  {P.}~\bibnamefont {Boccaccio}}, \bibinfo {author} {\bibfnamefont
  {F.}~\bibnamefont {Cavanna}},  \emph {et~al.},\ }\href
  {https://www.sciencedirect.com/science/article/pii/0168900294014051}
  {\bibfield  {journal} {\bibinfo  {journal} {Nucl. Instr. and Meth. A}\
  }\textbf {\bibinfo {volume} {355}},\ \bibinfo {pages} {660} (\bibinfo {year}
  {1995})}\BibitemShut {NoStop}%
\bibitem [{\citenamefont {Sorel}(2014)}]{sorel2014}%
  \BibitemOpen
  \bibfield  {author} {\bibinfo {author} {\bibfnamefont {M.}~\bibnamefont
  {Sorel}},\ }\href {https://doi.org/10.1088/1748-0221/9/10/p10002} {\bibfield
  {journal} {\bibinfo  {journal} {JINST}\ }\textbf {\bibinfo {volume} {9}},\
  \bibinfo {pages} {P10002} (\bibinfo {year} {2014})}\BibitemShut {NoStop}%
\bibitem [{\citenamefont {Sobel}(1968)}]{sobel}%
  \BibitemOpen
  \bibfield  {author} {\bibinfo {author} {\bibfnamefont {I.}~\bibnamefont
  {Sobel}},\ }\href@noop {} {\bibfield  {journal} {\bibinfo  {journal}
  {Presentation at Stanford A.I. Project}\ } (\bibinfo {year}
  {1968})}\BibitemShut {NoStop}%
\bibitem [{\citenamefont {Paszke}\ \emph {et~al.}(2019)\citenamefont {Paszke},
  \citenamefont {Gross}, \citenamefont {Massa} \emph {et~al.}}]{pytorch}%
  \BibitemOpen
  \bibfield  {author} {\bibinfo {author} {\bibfnamefont {A.}~\bibnamefont
  {Paszke}}, \bibinfo {author} {\bibfnamefont {S.}~\bibnamefont {Gross}},
  \bibinfo {author} {\bibfnamefont {F.}~\bibnamefont {Massa}},  \emph
  {et~al.},\ }in\ \href
  {http://papers.neurips.cc/paper/9015-pytorch-an-imperative-style-high-performance-deep-learning-library.pdf}
  {\emph {\bibinfo {booktitle} {Advances in Neural Information Processing
  Systems 32}}}\ (\bibinfo  {publisher} {Curran Associates, Inc.},\ \bibinfo
  {year} {2019})\ pp.\ \bibinfo {pages} {8024--8035}\BibitemShut {NoStop}%
\bibitem [{\citenamefont {James}(1998)}]{james1998}%
  \BibitemOpen
  \bibfield  {author} {\bibinfo {author} {\bibfnamefont {F.}~\bibnamefont
  {James}},\ }\href {https://cds.cern.ch/record/2296388} {\enquote {\bibinfo
  {title} {{MINUIT}: Function minimization and error analysis reference
  manual},}\ } (\bibinfo {year} {1998})\BibitemShut {NoStop}%
\end{thebibliography}%

\pagebreak
 
\end{document}